\documentclass[conference]{IEEEtran}
\IEEEoverridecommandlockouts

\usepackage{amssymb}
\usepackage{graphicx}
\usepackage{qtree}
\usepackage{circuitikz}
\usepackage{tikz}
\usepackage{mathtools}
\usepackage{tkz-berge}
\usepackage{booktabs}
\usepackage{hyperref}
\usepackage{subcaption}
\usepackage{dirtree}
\usepackage{booktabs}
\usepackage{multirow}

\usetikzlibrary{shapes,snakes}

\usepackage{amsthm}

\theoremstyle{definition}

\newtheorem{example}{Example}

\usepackage[linesnumbered,ruled,vlined]{algorithm2e}
\usepackage{algpseudocode}

\usetikzlibrary{shapes.gates.logic.US,trees,positioning,arrows}

\usepackage{lipsum}

\usepackage[latin1]{inputenc}
\usetikzlibrary{shapes,arrows}

\usepackage{pgfplots}
\pgfplotsset{width=8.3cm,compat=1.9}

\usepackage{pgfplotstable}
\usepackage{pgfplots}
\usetikzlibrary{patterns}
\pgfplotsset{compat=1.3}
\usepackage{xcolor}
\usepackage{csvsimple}

\newcommand{\new}{\textcolor{black}}
\usepackage[linesnumbered,ruled,vlined]{algorithm2e}
\usepackage{algpseudocode}

\begin{document}

\title{Cyber Situation Awareness Monitoring and Proactive Response for Enterprises on the Cloud}

\author{\IEEEauthorblockN{Hootan Alavizadeh\textsuperscript{1}, Hooman Alavizadeh\textsuperscript{2} and Julian Jang-Jaccard\textsuperscript{2}}
		\IEEEauthorblockA{\textsuperscript{1} Computer Engineering Department,\\ Imam Reza International University, Mashhah, Iran.\\
		Email: h.alavizadeh@imamreza.ac.ir\\
		\textsuperscript{2} School of Natural and Computational Sciences,\\ Massey University, Auckland, New Zealand.\\
		Email: \{h.alavizadeh,J.Jang-jaccard\}@massey.ac.nz}
}

 \maketitle

	\begin{abstract}
The cloud model allows many enterprises able to outsource computing resources at an affordable price without having to commit the expense upfront. Although the cloud providers are responsible for the security of the cloud, there are still many security concerns due to inherently complex model the cloud providers operate on (e.g.,multi-tenancy). In addition, the enterprises whose services have migrated into the cloud have a preference for their own cybersecurity situation awareness capability on top of the security mechanisms provided by the cloud providers. In this way, the enterprises can monitor the performance of the security offerings of the cloud and have a choice to decide and select potential response strategies more appropriate to the enterprise in the presence of the attack where the defense provided by the cloud doesn't work for them. However, some response strategies, such as Moving Target Defense (MTD) techniques shown to be effective to secure cloud, cannot be deployed by the enterprise themselves. In this paper, we propose a framework that enables better collaboration between enterprises and cloud providers. Our proposed framework, which offers more in-depth security analysis based on the set of most advanced security metrics, allows the security experts of the enterprise to obtain better situational awareness in the cloud. With better and more effective situation awareness of cloud security, our framework can support better decision making and further allows to deploy more appropriate threat responses to protect the outsourced resources. We also propose a secure protocol which can facilitate more secure communication between the enterprises and cloud provider. Using our proposed secure protocol, which is based on authentication and key exchange mechanism, the enterprises can send a secure request to the cloud provider to perform a selected defensive strategy.
	\end{abstract}

\section{Introduction}
Cloud computing is a powerful and affordable network paradigm providing the computational and storage requirements of individuals, enterprises, or governments. Many organizations decide to outsource their infrastructures and underlying services to the cloud to save untested up front expense. However, there  exist many security concerns for the third-party owned and operated the cloud environment where many details of the cloud operation are unknown for clients~\cite{bisong2011overview, zissis2012addressing}. 
The first issue is that cloud computing is a multi-tenant environment hosting various virtual machines (VMs) in the shared infrastructures, such as physical hosts (Servers). Such feature may enable adversaries to find vulnerabilities and launch various attacks (e.g., Side-channel attack~\cite{bates2012detecting,moon2015nomad}). The second problem is that cloud providers do not usually undertake security analysis and evaluation for the Software Defined Networks (SDN) or Virtual Networks (VN) created by customers on the cloud infrastructures and only rely on their own security mechanisms (e.g., firewalls and Intrusion Detection System (IDS)). 

\new{Situation Awareness (SA) systems can be used by cloud's customers to monitor their infrastructure in the cloud and gain awareness of the current and future security posture of their assets in the cloud. Although cloud providers have their own security mechanisms to protect the clients' assets, the organizations would need to have their own security SA platform including situation monitoring, situation modeling and analysis to improve their security posture. They may also need to have their own decision making and response selection to defend against possible threats.}



\begin{figure}[b]
	\centering
	\includegraphics[height=5.5cm]{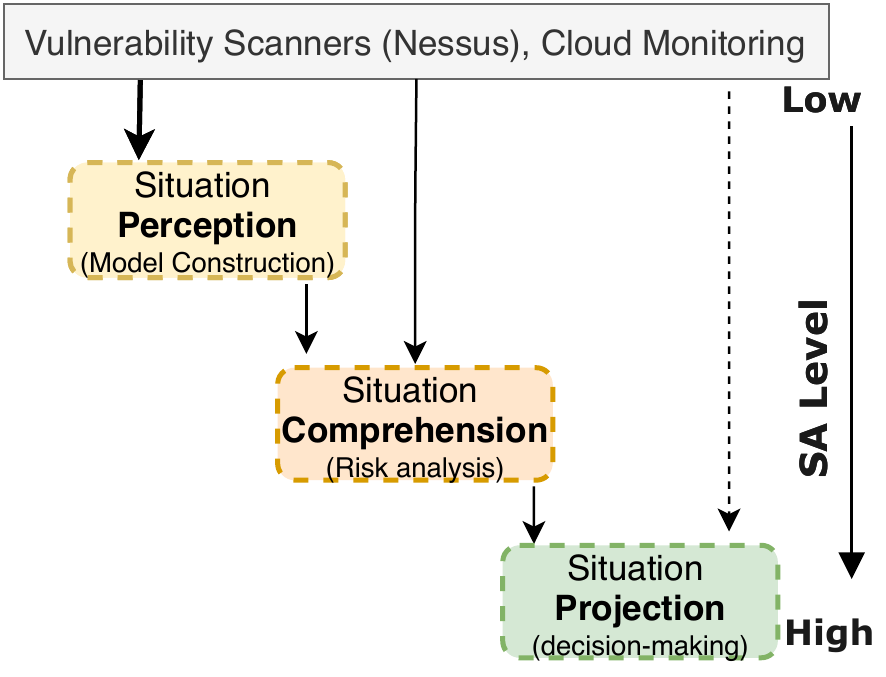}
	\caption{Situation Awareness phases and their relations}
	\label{fig:SA-levels}
\end{figure}

\new{Situation Awareness in cybersecurity includes the seminal aspects~\cite{endsley1988design}, as shown in Fig.~\ref{fig:SA-levels}: (i) Situation perception which involves awareness of the threats and vulnerabilities on the system which can provide the system administrator with a low level of understanding on the current situation. It includes data collection through various sources such as firewalls and vulnerability scanners, and so fourth. (ii) Situation comprehension which includes cloud risk assessment, attack and cost assessments, and so forth which provide a higher understanding or awareness of the current situation (security posture). (iii) Situation Projection including awareness of how the situation may further evolve and can further affect the system. This provides the highest level of situation awareness. It deals with more in depth analysis of the current situation and decision-making based on the impact analysis of the possible responses. It also deals with the effectiveness of the responses or countermeasures.}

Some Moving Target Defense (MTD) techniques (such as Virtual Machine Live Migration (VM-LM)) have been shown to be useful defensive strategies to secure the cloud computing as they can increase the security of the cloud by increasing the attackers time and effort through changing the attack   surface.~\cite{hong2016assessing,alavizadeh2017effective,alavizadeh2019effective}.  For instance, deploying VM-LM can improve cloud's security, decrease risk and attack success probability, and thwart VM co-residency attacks~\cite{alavizadeh2018evaluation,alavizadeh2019model}. However, VM-LM technique is usually restricted for cloud users and can only be performed by the cloud providers due to load balancing and security consideration of the cloud providers~\cite{alavizadeh2019An}. As VM-LM can not be performed directly by the clients, alternatively, the security experts of the organisations can send a request to the cloud providers to deploy the requested defensive strategy.

Proposing a secure platform to enable the security experts of the organizations to monitor and analyze the cloud security posture and obtain situation awareness themselves alongside with the cloud providers' security mechanism can increase the cloud's trustworthiness. Thus, the cooperation of the cloud providers and enterprises to secure the cloud environment can be beneficial for both parties. In this paper, we propose a framework with a secure protocol that enables organizations to monitor cloud and perform security analysis such as risk assessment for their network on the cloud and also select a defensive response strategy which can be either simply patching vulnerabilities or deploying proactive responses such as MTD techniques. However, deploying MTD techniques such as VM-LM approach may have various advantages: (i) mitigating the possibilities of co-residency problems such as side-channel attack for the Virtual Machines (VMs) located into the cloud. (ii) increasing the attack hardening due to changing the attack surface and make it unpredictable for the attackers, (iii) reducing the cloud risk values by adjusting the best MTD deployment scenario.
Thus, it is important to monitor the cloud security situation based on the vulnerabilities existing in the cloud to perceive the threats and then analyze the cloud's security posture (such as risk assessment) to obtain more comprehensive situation awareness of the cloud. Finally, make a decision about the best response plan which increases the security benefits and decreases the negative impacts on the cloud before choosing any MTD techniques to deploy. Fig.~\ref{fig:SA-levels} shows the situation awareness phases showing the SA levels. Selecting an appropriate response strategy can help the enterprises to reduce or hold the cloud risk on an acceptable threshold.

The main contributions of this paper are summarized as follows:

\renewcommand\labelitemii{$\bullet$}
\let\labelitemi\labelitemii

\begin{itemize}
	\item{\new{We propose a security situation awareness platform to monitor the assets in the cloud and discover the threats involved to gain the \textit{situation perception} and analyze the current situation of the cloud in terms of cloud risk for the enterprises migrated into the cloud to gain the \textit{situation comprehension}.}}
	\item{\new{We include the \textit{situation projection} and response in to the proposed SA system which enables the enterprises to perform further analysis on the current situation gained and make a decision about the available responses and the effects of each on the cloud by evaluating the defensive strategies like vulnerability patching or VM-LM;}}
	\item{Then, we propose a client-server based communication protocol including a key exchange scheme for setting up a secure communication between the enterprises' servers outside of the cloud and the cloud provider in order to deploy VM-LM as the main MTD technique on the cloud through sending a secure request to the cloud provider;}
	\item{We develop and validate our proposed protocol on a real private cloud.}
\end{itemize}

The rest of the paper is organized as follows.
In Section \ref{sec:pre}, we define the proposed approach including a brief explanation on required concepts, notations, and definitions. In Section \ref{sec:proposed}, the design and implementation of the proposed approach are given. Discussion and limitations of this study are given in Section \ref{sec:discussion}. In Section \ref{sec:RW}, the related work is summarized. Finally, we conclude the paper in Section \ref{sec:conclusion}.


\begin{figure*}[t]
	\centering
	\includegraphics[height=7cm, width=17.5cm]{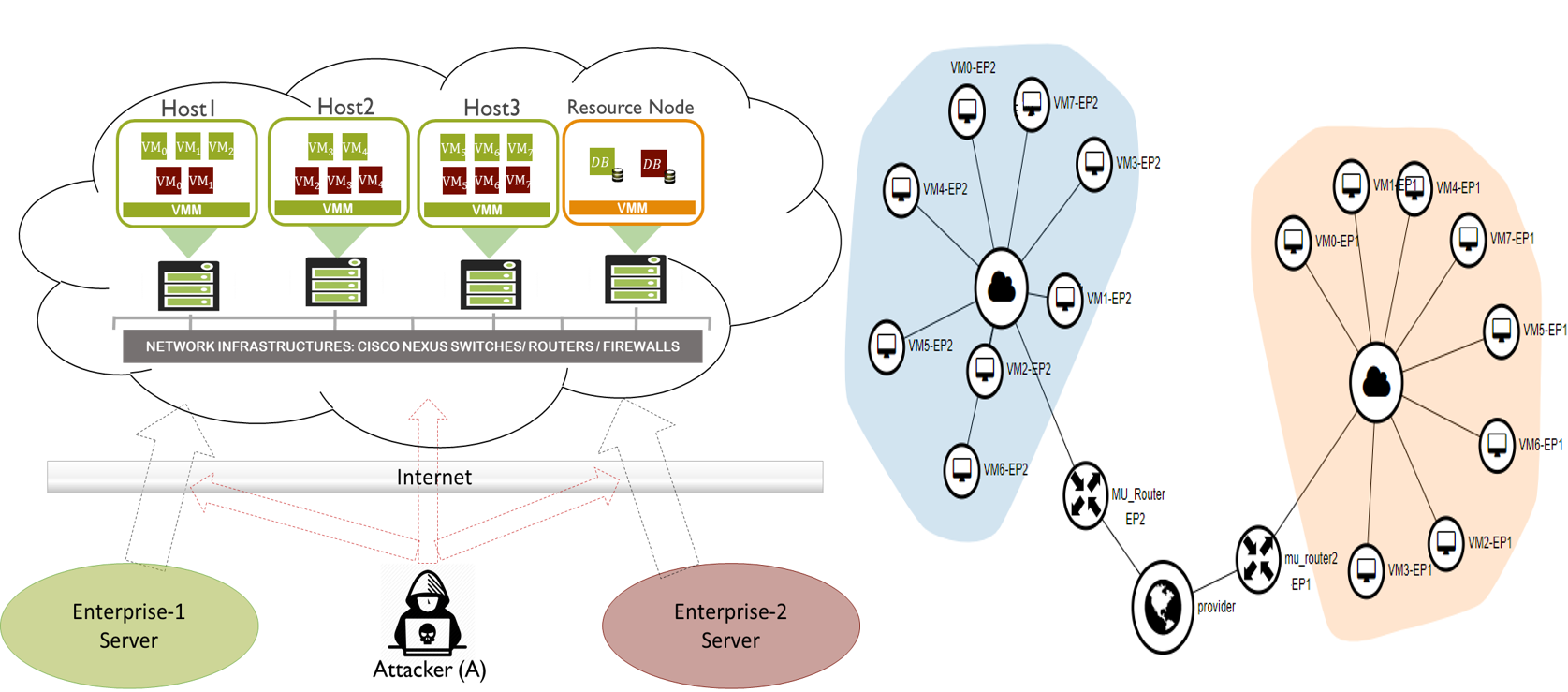}
	\caption{Running Example: Infrastructures and threat model and deployed model in the real cloud}
	\label{fig:running example}
\end{figure*}

\section{Proposed Approach}\label{sec:pre}
\subsection{Preliminaries}
In this section, we describe the related notations, concepts, and definitions which are used throughout this paper. We first define a running example as the main scenario for the migration of enterprises to the cloud. Then, we deploy the proposed approach on a real cloud environment~\cite{UniteCloud,he2016reverse}.

\begin{figure}[b]
	\centering
	\includegraphics[height=4.5cm]{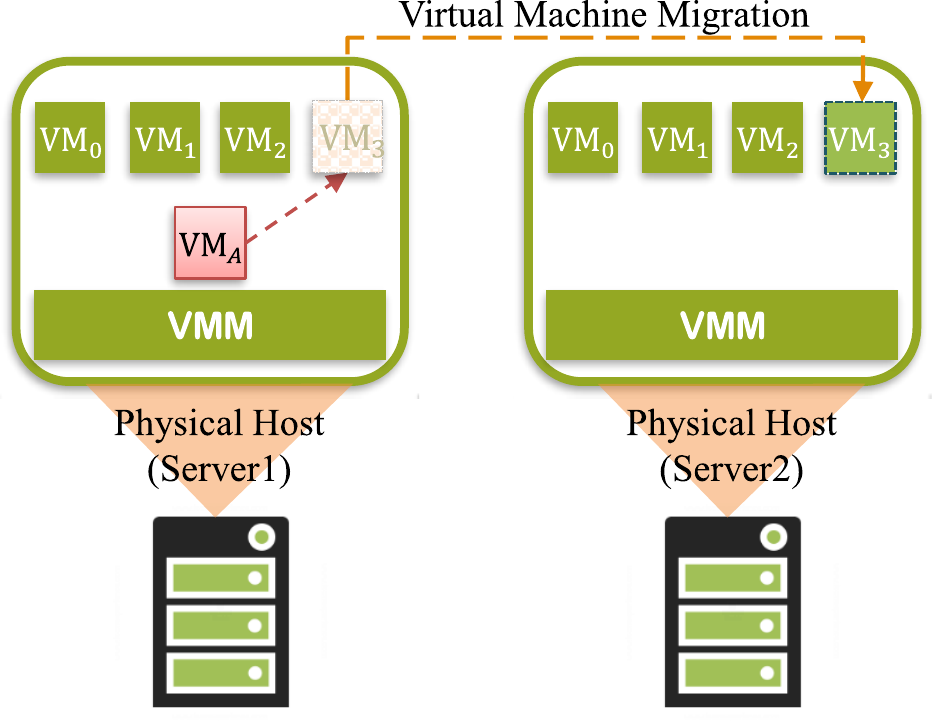}
	\caption{Co-residency and migration example}
	\label{fig:CO}
\end{figure} 

\subsection{Moving to Cloud Scenario}
Fig.~\ref{fig:running example} shows the running example scenario for the proposed approach based on the migrations of two independent enterprises entitled Enterprise-1 ($\text{EP}_1$) and Enterprise-2 ($\text{EP}_2$) that presume to reside within a private cloud. Those organisations decide to cut off the physical equipment and use a private cloud for accommodating their network infrastructures. Each organization has launched 8 VMs on the cloud together with a Database (DB) for creating a virtual network. We assume that the first 4 VMs use Windows10 instances and the rest are based on an Ubuntu operating system. The security experts of the organizations are responsible for the security of their infrastructures located on the cloud through the servers from outside of the cloud. The security experts need to perform the security monitoring and analysis (e.g. risk assessment) to gain security SA of the cloud, and perform a decision making and further to select a response for improving the security (e.g. by deploying MTD techniques). We implement the infrastructures of these two organizations on a real private cloud named UniteCloud~\cite{UniteCloud}.

\begin{figure}[b]
	\centering
	\includegraphics[height=3.25cm]{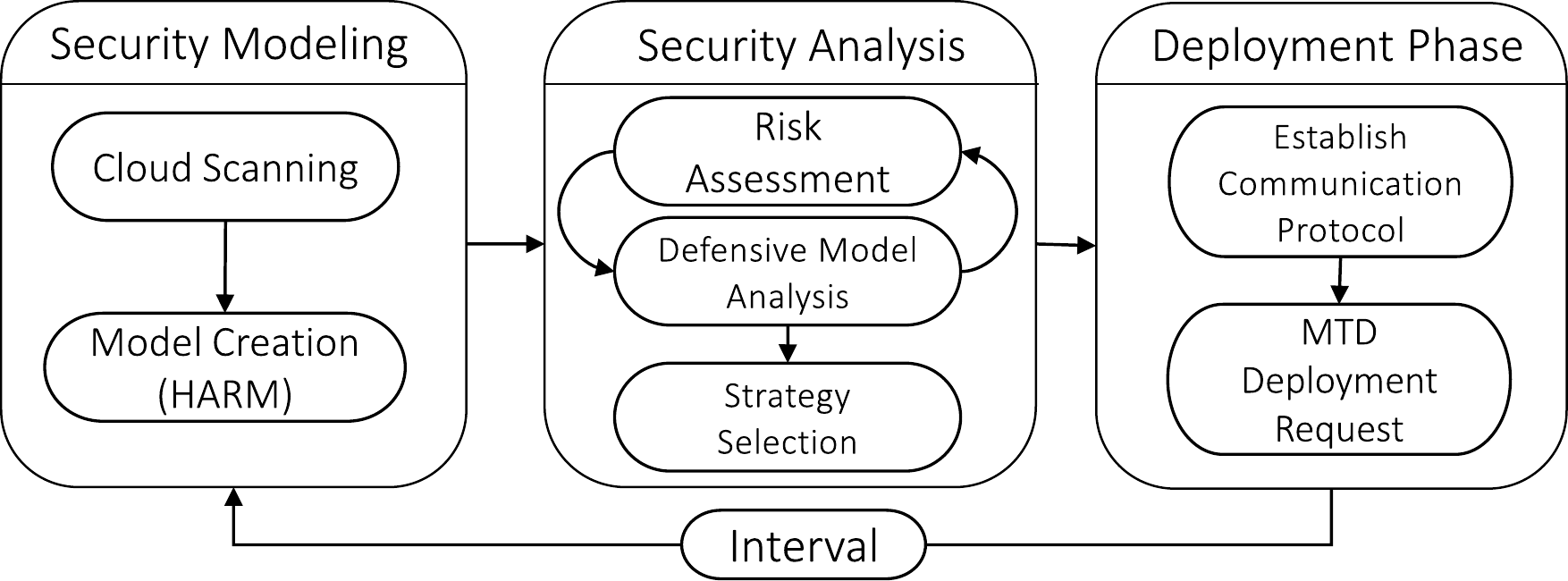}
	\caption{Situation Awareness Phases: Security Modeling, Analysis, and Deployment Phases}
	\label{fig:model}
\end{figure}

\subsection{Attack Models}
We assume that the attackers can launch the attacks as follows. (i) Exploiting the DB through the attack on the VN: the attacker can use the vulnerabilities existing on VM and launches an external attack. The attacker first launches a remote attack from the Internet to the VMs which are connected to the Internet. Then, the attacker can find an attack path and finally exploit the DB. (ii) Accessing a VM through side-channel: the attacker may try to reconnaissance a goal and create a VM in the same physical host with the target. Then, the attackers can benefit from the co-residency issues and access to the target VM, see Fig.~\ref{fig:CO}. (iii) Communication link: Attackers can attack the communication link between the enterprise servers located outside of the cloud and the cloud provider server on the cloud.

\subsection{Analysis and Defensive Model}
In order to thwart the possible attacks on the cloud infrastructure, the organizations can obtain the cloud security SA by performing data gathering regarding the possible vulnerabilities, and further risk assessment through the security analysis from their servers located outside of the cloud and, finally, select an appropriate threat response. In this paper, we utilized VM-LM as the main proactive response.

\begin{figure*}[t]
	\centering
	\begin{subfigure}{0.50\textwidth}
		\centering
		\includegraphics[height=8cm, width=8cm]{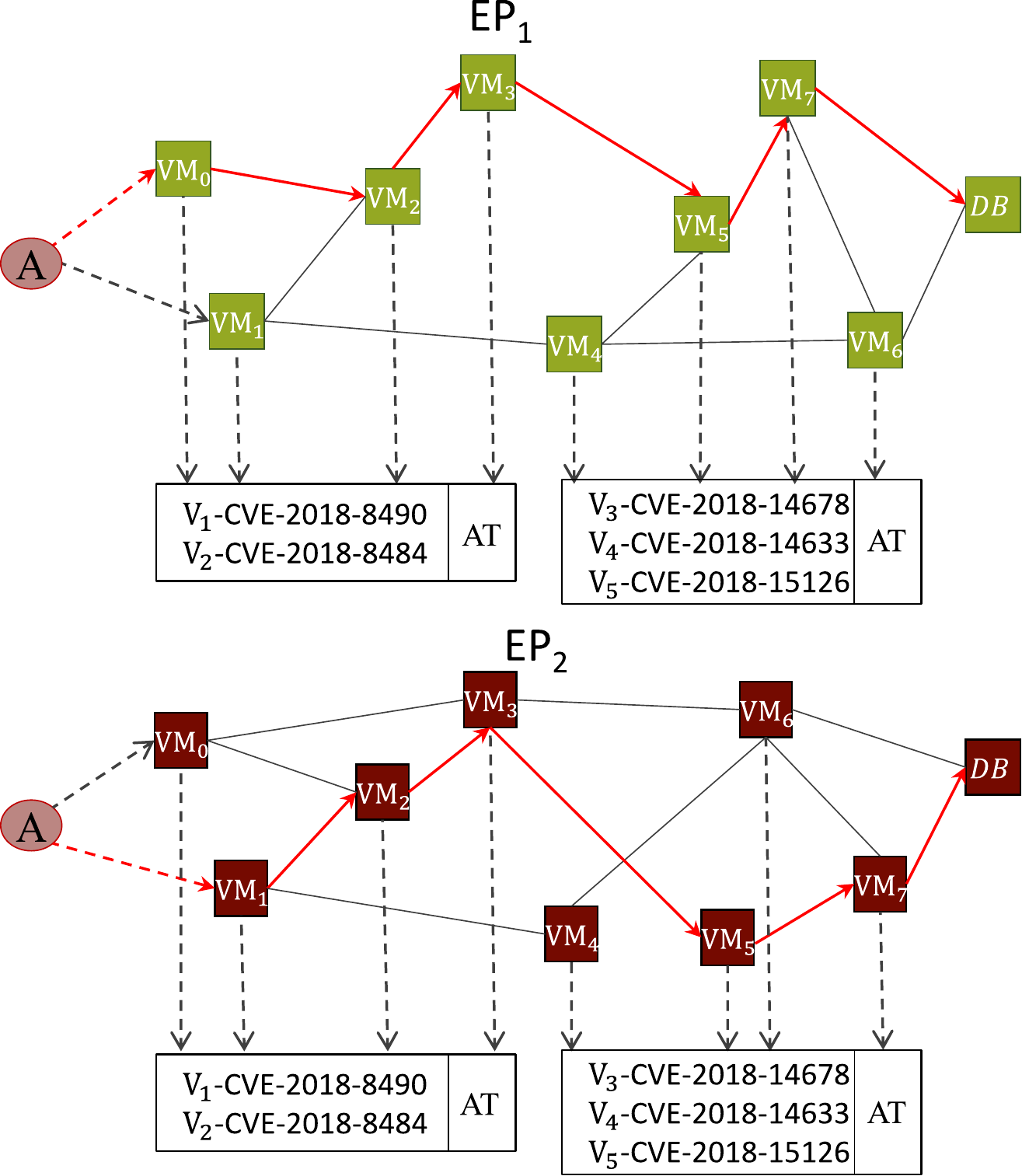}
		\caption{}
		\label{fig:HARMs}
	\end{subfigure}
	\begin{subfigure}{0.45\textwidth}
	\centering
	\includegraphics[height=8cm]{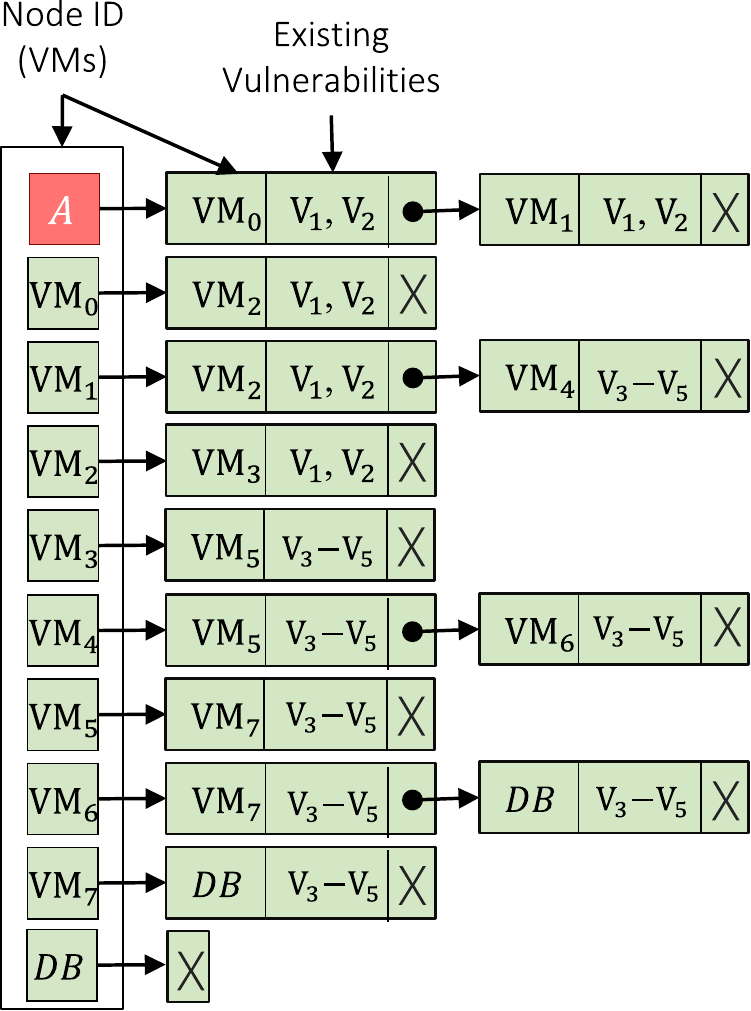}
	\caption{}
	\label{fig:adj}
	\end{subfigure}
	\caption{Generated HARM for $\text{EP}_1$ and $\text{EP}_2$ including OS Vulnerabilities on each VM. (a) The Upper and Lower layers of HARM. (b) An adjacency lists for $\text{EP}_1$ graph. }
	\label{fig:HARM}
\end{figure*}

VM-LM can be deployed one time or set periodically (a Time-based MTD~\cite{alavizadeh2019effective}) as the main MTD technique by aiming to decrease the attacker's success rate on finding a target on a physical machine and lower the chance of attacker to locate a malicious VM in the same machine with the target, see Fig.~\ref{fig:CO}. It also can reduce the cloud risk and increase the security by changing the attack surface and make confusion for the attackers~\cite{alavizadeh2019model}. The selected MTD technique should be chosen in a way that it holds the system risk in an acceptable range. Time-based MTD approaches deploy MTD technique (in here, VM-LM) after passing a certain time interval~\cite{Cai16}. More frequency of deploying a VM-LM operation provides more complexity and difficulties for the attacker to reconnaissance and co-locate a malicious VM on the same physical host of the target. Moreover, as the organizations perform the security analysis and cloud provider deploys the requested MTD strategies, the collaboration and communication between organizations and cloud providers should be secure enough. We propose a security situation awareness framework consisting of security modeling (situation perception), security analysis (situation comprehension), and decision making and deployment phase (situation projection), as shown in Fig.~\ref{fig:model}. The security modeling phase includes information gathering and constructing security models based on the obtained information. Then, the security analysis (comprehension) phase computes some security metrics such as the cloud risk to evaluate the overall cloud security, Return on Attack (RoA) to evaluate security from the attacker's perspective, and Mean of Attack Path Length (MAPL) to observe the attack hardening~\cite{alavizadeh2019model}. Obtaining the current situation of the cloud, the security experts of the organisations can evaluate different response scenarios to find the best response to deploy.
As stated earlier, VM-LM strategy is used by aiming to make the attack harder and more confusing.
However, VM-LM can be deployed in various VMs in the cloud. Thus, it is the responsibility in the security projection phase to find out which VM needs to be selected for deploying VM-LM so that it has the best effect on the security while it makes the lower future impact on the system such as improving or deteriorating the cloud risk. Thus, the defensive model should analyze the effects of each VM-LM scenario on the cloud risk so that the cloud risk can be preserved on an acceptable level after each VM-LM deployment. Finally, a secure request from the enterprises to the cloud provider's server should be sent including the selected VM-LM strategies for deployment on the cloud. More technical details for the proposed framework including the design and implementation are given in the following section.

\section{Design and Implementation}\label{sec:proposed}
We utilize the following concepts and tools to develop and implement a secure cloud analysis protocol. We utilize Nessus~\cite{nessus} which is a vulnerability scanning tool, Common Vulnerabilities Scanning Systems (CVSS), Hierarchical Attack Representation Model (HARM)~\cite{hong2012harms}, OpenStack APIs, .NET Core, Data-Driven Documents JavaScript (D3.js).

\subsection{\new{Situation Perception using Security modeling}}
Security modeling is the first phase of the cloud security framework which can provide a preliminary perception of the model and its vulnerabilities. 
The servers at the enterprise from outside of the cloud are responsible for monitoring and analysing their infrastructures on the cloud. This phase consists of two steps: (1) cloud scanning that includes network and vulnerability scanning, and (2) model creation using HARM. The details of generating HARM is given in~\cite{alavizadeh2019An}. Cloud scanning phase uses OpenStack APIs to collect the required information such as the number of VMs, Hosts, Connectivity of VMs, etc. Moreover, it captures the vulnerabilities existing on each VMs through Nessus tool and APIs to automatically obtain the scan reports~\cite{alavizadeh2019An}. The main results of the first phase are a set of VMs and the reachabilities of them, and a set of vulnerabilities (V) existing on each VM. In the next step, HARM can be constructed using the obtained information. In this paper, we only extract and parse OS vulnerabilities from vulnerability scan report for simplicity. However, other vulnerabilities, such as services, applications, etc., can also be extracted and incorporated into the model. HARM consists of two layers which captures VMs and their reachabilities on the upper layer using Attack Graph (AG) and the existing vulnerabilities on the lower layer using Attack Tree (AT)~\cite{alavizadeh2019model}. The entry points of the cloud are the VMs connected to the Internet and the target is the DB. Both entry points and target are captured in the upper layer of the HARM, see Fig.~\ref{fig:HARMs}. The generated HARM for the running example is represented as adjacency lists including an additional node denoted as $A$ for the attacker. For instance, Fig.~\ref{fig:adj} shows the adjacency lists of the upper layer of HARM for $\text{EP}_1$.

 


\begin{figure}[t]
	\centering
	\includegraphics[height=4.25cm]{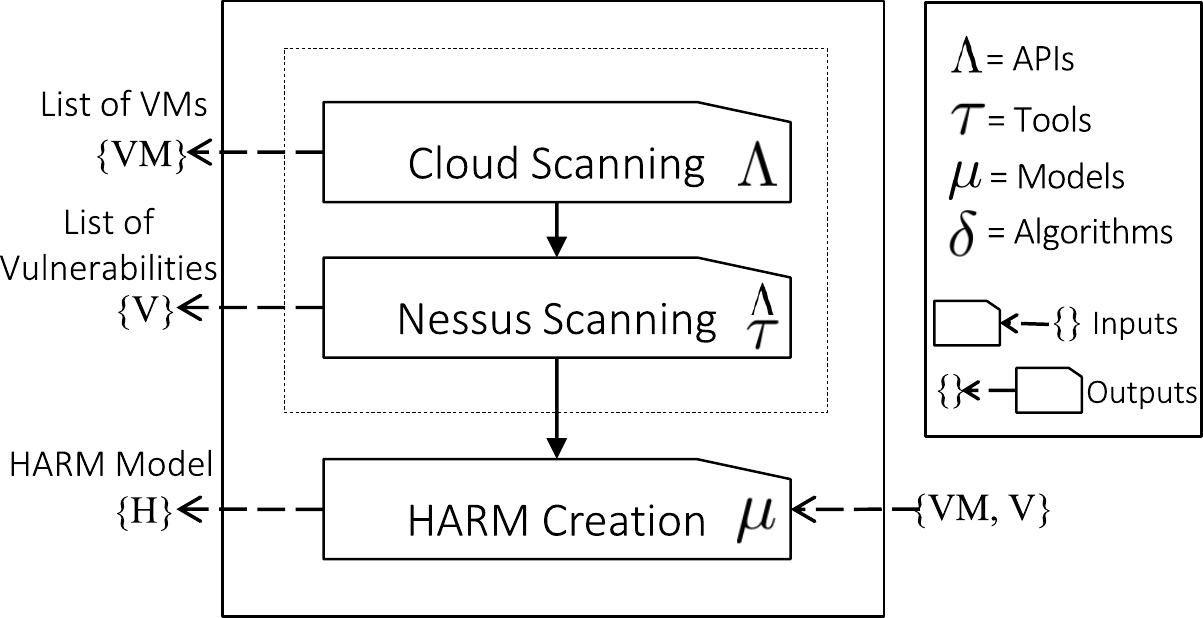}
	\caption{Situation Perception: Security modeling phase procedures}
	\label{fig:phase1}
\end{figure}
\begin{figure}[b]
	\centering
	\includegraphics[height=5.5cm, width=8.7cm]{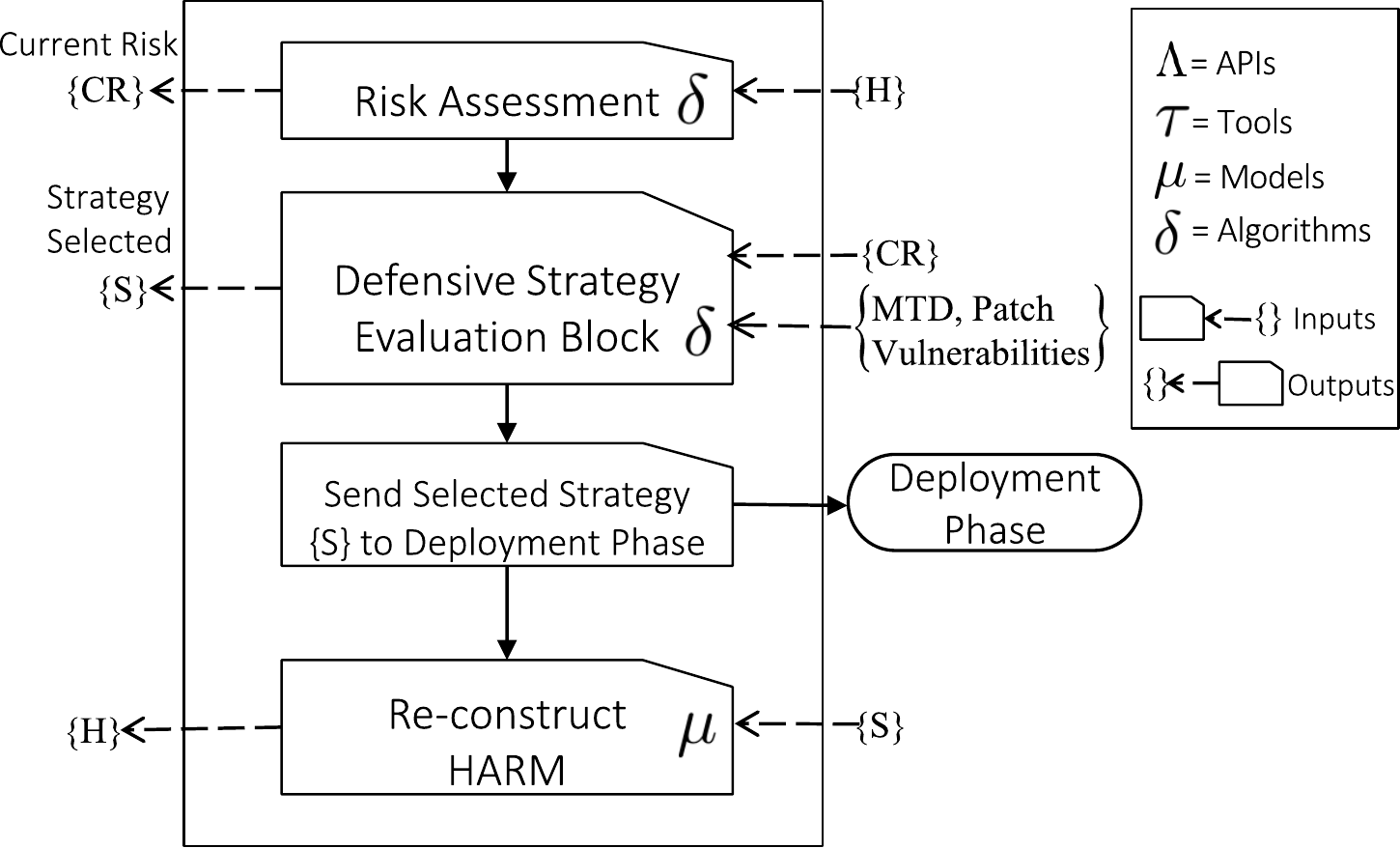}
	\caption{Situation Comprehension and Projection: Security analysis and strategy selection phase}
	\label{fig:phase2}
\end{figure}
\subsection{\new{Situation Comprehension and Projection}}\label{subsec:SA}
In this paper, we use risk assessment as one the main security analysis criteria which can calculate the risk value of the enterprises' infrastructure on the cloud. Risk assessment is actually a systematic and analytical process to consider the likelihood that a threat can endanger an asset~\cite{dimopoulos2004approaches}. In this phase, the result of the generated HARM (denoted as H) is used as an input for the security analysis and defensive strategy selection procedures. HARM uses an AG in the upper layer and a AT in the lower layer. Vulnerabilities are captured on the lower layer. The vulnerabilities include three main metrics based on the National Vulnerability Database (NVD) \cite{mell2006common}: BaseScore (BS), Exploitability (E), and Impact (I), and Attack Cost (AC).
For instance, there are two vulnerabilities named as CVE-2018-8490 ($\text{V}_1$) with ($E=0.17$, $I=6$) and CVE-2018-8484 ($\text{V}_2$) with ($E=0.18$, $I=5.9$) for Windows 10, and three vulnerabilities for Ubuntu entitled as CVE-2018-14678 ($\text{V}_3$) with ($E=0.18$, $I=5.9$), CVE-2018-14633 ($\text{V}_4$) with ($E=0.22$, $I=4.7$), and CVE-2018-15126 ($\text{V}_5$) with ($E=0.22$, $I=5.9$). The attacker can penetrate into a VM by exploiting only one of the vulnerabilities, thus, for the AT we use an OR-gate. The risk of exploiting a VM (R($\text{VM}_i$)) can be calculated as the $R(\text{VM}_i)=E_i\times I_i$, as in~\cite{hong2016assessing}. If a VM includes more than one vulnerability, then the metrics of the vulnerability having higher severity is used (OR-gate). For example, the severity of $\text{V}_4$ is as $4.7\times 0.22=1.03$ and this rate for $\text{V}_5$ is almost $1.3$. Thus, $\text{V}_5$ is used for calculation of risk for Ubuntu. The following example shows the risk of exploiting a possible path (highlighted in Fig.~\ref{fig:HARMs}) from the attacker to the DB for $\text{EP}_1$.
	\begin{example}\label{ap}
	The path risk (PR) value for a single path (ap) can be computed as sum of the risk associated with each VM on the path which is as  $R(\text{VM}_0)+R(\text{VM}_2)+R(\text{VM}_3)+R(\text{VM}_5)+R(\text{VM}_7)+R(\text{DB})= 5.9*0.18+ 5.9*0.18+ 5.9*0.18+5.9*0.22+ 5.9*0.22+ 5.9*0.22=7.08$.
	\end{example}

\begin{algorithm}[t]
	\SetAlgoLined
	\DontPrintSemicolon
	\KwData{$\{\text{H}\}$ \footnotesize \tcc*[f]{H: HARM}}
	\KwResult{$\{\text{CR}\}$ \footnotesize \tcc*[f]{CR: Cloud Risk}}
	\Begin{
		$VM\leftarrow$Get-VM-List(H)\;
		$AP\leftarrow$Get-All-Attack-Paths(H)\;
		\ForEach {$ap \in AP$}{
			\ForEach {$VM_i \in ap$}{
				{$\text{CR} = \text{CR} +R(VM_i)$ \footnotesize \tcc*[f]{$R(VM_i)=E_i\times I_i$}}
			}
		}
		\KwRet{{CR} \footnotesize \tcc*[f]{Return Total Cloud Risk}}
	}
	\caption{Risk Assessment Procedure (RA): Computation of risk value}
	\label{alg:RA}
\end{algorithm}
	
\new{Moreover, we leverage two more security metrics which helps to gain situation awareness in terms of attacker's perspective such as Return on the Attack (RoA) and Mean of Attack Path Length (MAPL). The benefit of exploiting the vulnerabilities on a VM by an attacker is defined as RoA which can be obtained as $RoA_i=R(VM_i)/(AC_i)$ as in~\cite{alavizadeh2019model}. Moreover, the existence of more number of attack paths ($p$) in the cloud indicates less security as the attacker can exploit alternative paths to reach the target. Thus, the higher MAPL indicates less security in the network. Equation~\ref{eq:mapl} shows the calculation for MAPL based on the HARM represented in Fig.~\ref{fig:HARM}.}, where $ap$ denoted a single attack path and $AP$ shows a set of all possible attack paths in the cloud.

\begin{equation}\label{eq:mapl}
\text{MAPL}=\frac{\sum_{ap \in AP} |ap|}{p}
\end{equation}

Algorithm~\ref{alg:RA} shows the procedure of risk assessment in order to calculate the overall cloud risk based on the generated HARM. Once the cloud risk value is calculated, the cloud security framework follows the next step to select a response strategy in order to defend against the possible attacks. However, the selected defensive strategy may vary the cloud risk value. Thus, the strategy should be wisely chosen so that it holds the cloud risk on an acceptable threshold. 
This movement confuses the attacker, as the attacker needs to spend more time and effort to find out the new location of the target and place his malicious VM on that physical server. Migration of VMs may change the VM connectivity and affect the upper layer of HARM. This procedure considers the migration of each single VM, and then computes the cloud risk again repeatedly. The result of this step is a VM-LM strategy which has the best results in terms of decreasing or keeping the cloud risk value at an appropriate level which can be determined by security experts. The defensive strategy evaluation and selection are given in Algorithm~\ref{alg:DSES}. Then, the selected VM-LM strategy having the best effect on the risk value (denoted by \textit{S}) is prepared to send to the cloud provider's server in the cloud for deployment on the real cloud. The protocol and process of establishing secure communication and sending the VM-LM request will be discussed on the deployment phase in the following section. Once the cloud provider acknowledges the organization that VM-LM is successfully deployed, the HARM is updated for the next iterations. Although the main goal of this paper is to use VM-LM as the main defensive strategy against cyber attacks on the cloud, we also consider the utilization of patching vulnerabilities for the OS vulnerabilities to compare the VM-LM which is a proactive MTD technique with a usual response technique which only remove the detected vulnerability.

\begin{algorithm}[t]
	\SetAlgoLined
	\DontPrintSemicolon
	\KwData{$\{\text{CR}\}$ \footnotesize \tcc*[f]{CR: Cloud Risk obtained from risk assessment step}}
	\KwData{$countermeasures$= \{MTD, Patching\} \footnotesize \tcc*[f]{MTD: VM-LM}}
	\KwResult{$\{\text{S}\}$ \footnotesize \tcc*[f]{S: The selected VM-LM Strategy}}
	\Begin{
		$VM\leftarrow$Get-VM-List(H)\;
		$AP\leftarrow$Get-All-Attack-Paths(H)\;
		\ForEach {$vm_i \in VM$}{
			{$\text{H}'\leftarrow$~MTD($vm_i$,$\text{H}$) \footnotesize \tcc*[f]{Logically deploy VM-LM in H and returns new HARM $\text{H}'$}\;}
			{$\text{CR}'\leftarrow \text{RA}(\text{H}')$ \footnotesize \tcc*[f]{$\text{CR}'$: the new cloud risk value after VL-LM on $vm_i$}\;}
			{Add $\text{CR}'$ into $Z_{i}$ \footnotesize \tcc*[f]{$Z$: CR values list}\;}
		}
		{$i\leftarrow$ Select\big([$\min(Z)$]\big)\footnotesize \tcc*[f]{select a index of $Z$ with lowest value}\;}
		{S$=$(VM-LM, $VM_i$) \footnotesize \tcc*[f]{S is a tuple}\;}
		\KwRet{{S} \footnotesize \tcc*[f]{Return the VM-LM strategy on a selected VM}}
	}
	\caption{Defensive Strategy Evaluation and Selection}
	\label{alg:DSES}
\end{algorithm}

\begin{table}[t]
	\centering
	\caption{The percentages of changes on cloud risk resulting from security analysis phase for $\text{EP}_1$ and $\text{EP}_2$ through considering VM-LM and Patching vulnerabilities (selected strategy is denoted by ($\checkmark$)).}
	\label{TcombI}
	\begin{tabular}{@{}p{0.45cm}lp{0.97cm}llll@{}}
		\toprule
		\multirow{2}{*}{VM} & \multicolumn{3}{l}{\% of Changes ($\text{EP}_1$)}      & \multicolumn{3}{l}{\% of Changes ($\text{EP}_2$)} \\ \cmidrule(l){2-7} 
		& Patching & VM-LM & \textit{S}                    & Patching & VM-LM & \textit{S}        \\ \midrule
		$VM_0$ & -1.65\%     & -27.63\% & \multicolumn{1}{l|}{$\times$}    & -4.01\%     & -30.81\%     & $\times$    \\ 
		$VM_1$ & -4.94\%     & -20.05\% & \multicolumn{1}{l|}{$\times$}    & -3.00\%     & -16.82\%     &$\times$     \\ 
		$VM_2$ & -3.29\%     & -31.42\% & \multicolumn{1}{l|}{$\times$}    & -4.01\%     & -35.42\%     & $\times$    \\ 
		$VM_3$ & -3.29\%     & -7.58\% & \multicolumn{1}{l|}{$\times$}    & -6.01\%     & -49.41\%     & $\times$    \\ 
		$VM_4$ & -4.99\%     & -29.33\% & \multicolumn{1}{l|}{$\times$}    & -1.52\%     & -26.05\%     &$\times$     \\ 
		$VM_5$ & -7.49\%     & -33.98\% & \multicolumn{1}{l|}{$\times$}    & -4.56\%     & -12.20\%     &$\times$    \\ 
		$VM_6$ & -6.24\%     & -42.84\%  & \multicolumn{1}{l|}{$\times$}    & -8.36\%     & -45.57\%     &$\times$     \\ 
		$VM_7$ & -8.74\%     & -54.64\% & \multicolumn{1}{l|}{$\checkmark$}    & -7.60\%     & -54.95\%     & $\checkmark$   \\ \bottomrule
	\end{tabular}
\end{table}

\begin{figure*}[t]
	\begin{subfigure}{0.49\textwidth}
		\centering
		\includegraphics[height=7.5cm]{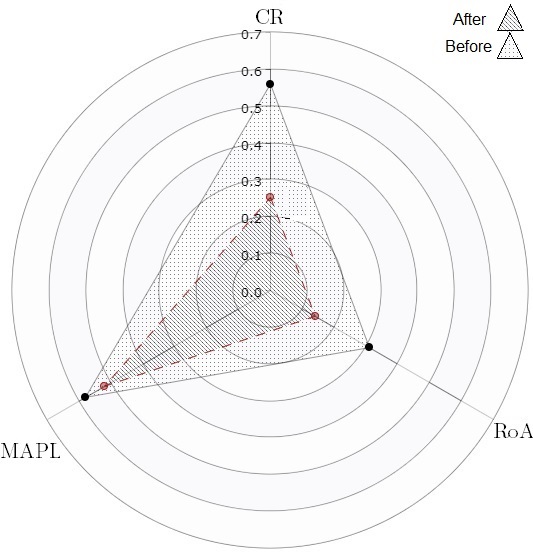}
		\caption{}
		\label{fig:EP1}
	\end{subfigure}
	\begin{subfigure}{0.49\textwidth}
		\centering
		\includegraphics[height=7.5cm]{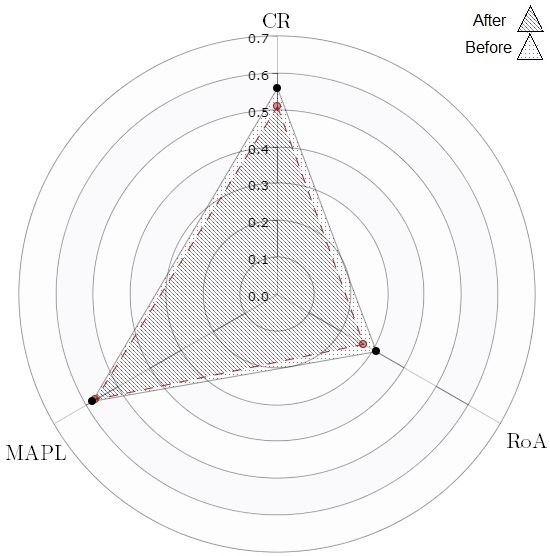}
		\caption{}
		\label{fig:EP2}
	\end{subfigure}
	\caption{Radar charts demonstrating the situation awareness of cloud security posture based on CR, RoA, MAPL metrics after (a) deploying VM-LM techniques and (b) patching vulnerabilities.}
	\label{fig:EP-Risks}
\end{figure*}

We then compare the results obtained from the patching the OS vulnerabilities and the VM-LM technique, as tabulated in Table~\ref{TcombI}. We analyze the results of deploying both VM-LM and patching vulnerabilities to all VMs on the cloud. The results show that VM-LM technique surpasses patching vulnerabilities due to the following reasons: (i) The VM-LM can provide a higher improvement on the cloud risk level, for example, in the best scenario for $\text{EP}_1$, the patching of the vulnerabilities of $\text{VM}_7$ decreases the cloud risk value by about 8.8\%, while this rate for VM-LM is about 54.6\%. Moreover, almost the same rates are observed for after analysis of $\text{EP}_2$, reported in the Table~\ref{TcombI}. (ii) Patching OS vulnerabilities cannot address the VM co-residency issues, as the malicious VM is still on the same physical machine of the victim. (iii) as stated earlier, all vulnerabilities are non-patchable, or difficult to patch. Accordingly, Fig.~\ref{fig:EP-Risks} represents a radar charts capturing the situation awareness of the cloud for $EP_1$ based on the security posture obtained from calculation of CR, RoA, MAPL metrics after deploying VM-LM techniques and patching vulnerabilities. The results show that deploying VM-LM yields better security posture in comparison with patching vulnerability as it decreases all values for RoA, CR, and MAPL.  
However, it is also important for the organizations to always identify and patch the patchable vulnerabilities alongside with any other defensive strategy like MTD. 

\subsection{Deployment Phase and Response}

\begin{table}[b]
	\centering
	\caption{Notations used in the paper}
	\label{notations}
	\begin{tabular}{@{}p{1.20cm}p{6.5cm}@{}}
		\toprule
		\textbf{Notations}   & \multicolumn{1}{c}{\textbf{Descriptions}}                     \\ \midrule
		$\hbar(x)$         	 & Message digestion function of x (MD5 hash function)        \\
		$\kappa_\rho$                & Public key (Asymmetric)                                       \\
		$\kappa_\gamma$                & Private key (Asymmetric)                                      \\
		$\kappa_{\rho-x}$            & Public key of an entity denoted by 'x'                        \\
		$\kappa_{\gamma-x}$            & Private key of an entity denoted by 'x'                       \\
		$E_\rho$                & Asymmetric encryption with a public key                       \\
		$E_\gamma$                & Asymmetric encryption with a private key                      \\
		$E_{\rho-x}$            & Asymmetric encryption with the public key of 'x'                  \\
		$E_{\gamma-x}$            & Asymmetric encryption with the private key of 'x'                 \\
		$\kappa_{\text{shared}}$         & Symmetric shared key for both parties                                         \\
		$E_{\kappa-\text{shared}}$       & Symmetric encryption by the shared key                        \\\bottomrule
	\end{tabular}
\end{table}
In this phase, the selected MTD technique resulting from the security analysis phase, given in Section~\ref{subsec:SA}, should be sent to the cloud provider for real deployment on the cloud's infrastructure. The main reason behind this request is the cloud constraints defined by cloud providers. Some countermeasures like patching or removing vulnerabilities (like patching OS and Software vulnerabilities, not hypervisors vulnerabilities) can be undertaken by the organizations themselves, but most of the cloud providers do not allow the customers to deploy the VM-LM strategy on the cloud. Alternatively, cloud providers can receive the request from the clouds' customers and deploy the VM-LM technique. In this section, we propose a protocol in order to establish secure communication between cloud providers and the organizations' servers. Then, organizations can securely send their VM-LM request to the cloud provider's servers. Authenticating the users, the cloud provider can deploy the requested VM-LM for authorized organizations. The defined protocol includes a key exchange scheme for establishing the connection.
The communication between the enterprises' servers and the cloud provider's server undergoes the following steps. (i) Both parties generate the public and private keys such that the public key of the cloud provider (CP) is known by organizations migrated into the cloud. Moreover, once an organization migrate to the cloud, it receives a secret enterprise code (EP-code). (ii) The enterprises (EP) prepare a request message including the EP-code and the organization's public key, then the message is encrypted by the cloud provider's public key and sent to the cloud provider's server, see Equation~\ref{eq:req}. Table~\ref{notations} shows some notations and descriptions used in this section.

\begin{equation}\label{eq:req}
\text{Registration~Request} = E_{\rho-\textsc{cp}}[\text{\textsc{ep}-\textit{code}} + \kappa_{\rho-\textsc{ep}}]
\end{equation}

Next, once the cloud server receives the request message and perform the authentication process, it obtains the user's public key included in the request message and creates a symmetric shared key ($\kappa_{\text{shared}}$) for the organization. Then, the server creates a reply message and replies it to the organization as Equation~\ref{eq:rep}.

\begin{equation}\label{eq:rep}
\text{Reply~Message} = E_{\rho-\textsc{ep}}[\kappa_{\text{shared}}]+\textsc{ack}+
\end{equation}
$$
E_{\gamma-\textsc{cp}}[nonce+\hbar(\textit{all fields})]
$$

Then, once the organization receives the reply message, it performs the authentication process using decryption of the third part of the reply message using the cloud provider's public key and comparing the fields. The organization obtains the shared key and acknowledgment for the registration. Later on, the communication between the organization and the cloud provider is based on the shared key. The messages for further communication from client to cloud are as Equation~\ref{eq:further}.

\begin{equation}\label{eq:further}
\text{Further~Message} = E_{\kappa\text{-shared}}[\textit{S}]+
\end{equation}
$$
E_{\gamma-\textsc{ep}}[nonce+\hbar(\textit{all fields})]
$$

Authenticating the message, the cloud provider's server uses the shared key to decrypt the message containing the requested strategy (denoted by \textit{S} in Equation~\ref{eq:further}), and then deploys the requested VM-LM strategy on the cloud. Finally, the cloud provider acknowledges the organization about the success or failure of the action.

\section{Discussion and Limitations}\label{sec:discussion}
Security situation awareness monitoring capabilities play an important role for finding the security holes on the cloud environment and making decision about the appropriate defensive responses. We propose a secure protocol which can help the organizations to have their own security monitoring and risk assessment for their assets on the cloud and deploy the defensive strategies on the cloud from the servers which are located outside of the cloud.
When the new vulnerabilities and threats are discovered, the key steps for the organizations are to assess the risk, then establish appropriate security countermeasures like patching the newly founded vulnerabilities or deploying MTD techniques to decrease the threats. 
We implement our framework in the real private cloud and analyze the cloud risk from outside of the cloud. The experimental results of the first round of running the protocol is reported in Fig.~\ref{fig:EP-Risks} and Table~\ref{TcombI}. The results compare the patching vulnerabilities and VM-LM as the defensive techniques and demonstrate that deploying VM-LM is more effective than patching vulnerabilities based on the cloud risk values together with RoA and MAPL which are metrics from the attacker's perspective. Moreover, patching vulnerabilities cannot avoid co-residency issues while VM-LM can address co-residency problems~\cite{zhang2012incentive}.

However, the communication portion of the proposed protocol provides the main security services ensuring the confidentiality and integrity of the messages, and also authentication of parties. The adversary cannot attack the communication protocol to read, forge, or alter the request messages including the VM-LM strategy because the messages are encrypted by a symmetric shared key together with a hash value of all messages' fields signed by the enterprises' private key. The value of nonce is also added to the signed portion of the message to avoid the reply attack. We also include a key exchange scheme on the protocol in order to securely exchange public key of the organizations to the cloud and receive the shared key from the cloud provider (as in Equations~\ref{eq:req} and \ref{eq:rep}). 

\textbf{Limitations.} The proposed protocol for deploy the VM-LM can be done periodically, but in this paper we only evaluate one round of response deployment. Moreover, VM-LM can also be adopted based on the responses to the events and security alerts t make it more promising defensive technique. For example, once an intrusion is detected on a physical server using the Intrusion Detection Systems (IDS), then an event for triggering VM-LM operation can be raised. However, we will further consider event-based VM-LM deployment in our future work.


\section{Related Work}\label{sec:RW}
Cloud security problems have been studying in various studies~\cite{bates2012detecting, han2017using, varadarajan2015placement}. Mreover, the application of situation awareness in the cloud also have been widely studied~\cite{chen2016cloud,yu2013cloud,zhang2018network}. In \cite{chen2016cloud}, the authors proposed a cloud computing based network monitoring and threat detection system to secure the critical infrastructure of the cloud using monitoring agents, cloud infrastructure, and an operation center. 
In another study~\cite{yu2013cloud}, the authors proposed a cloud computing based architecture for conducting cyber space situation awareness and leveraged the cloud infrastructure with a cost-effective data storage and investigated efficient threat detection techniques. Moreover, In \cite{zhang2018network}, the authors proposed a situational awareness method in cloud computing environment using a security analysis node named target virtual machine (TVM) and evaluated the impact of attack behavior on TVM by virtual machine introspection (VMI). However, the proposed methods have not considered situation comprehension and projection to make a decision and deploy an effective response to enhance the security. Moreover, most of the proposed methods have not been implemented on real cloud environment.

In~\cite{varadarajan2015placement}, authors investigated on the multi-tenancy problem and co-residency attacks, they reported that there is a high chance for attackers to find and locate their VMs into the same physical server of the victim even in the public cloud with various datacenters and physical servers. Moreover, a number of side channels have been explored \cite{hlavacs2011energy,wu2011identification} in order to transfer sensitive information between VMs, which is prohibited by security policies. 
However, most of the existing research work on theoretical aspects of the cloud security problems.

\textit{Li et al}.~\cite{li2012improving} proposed a virtual machine replacement strategy based on the security risk on the cloud by considering migration time and computing costs, but they didn't consider the security analysis using the security models and also their strategy is not optimal in terms of migration. 

Cloud security analysis through considering MTD techniques are also investigated by researchers~\cite{hong2019systematic, alavizadeh2018evaluation,alavizadeh2018comprehensive}. In~\cite{hong2016assessing}, the authors introduced three main MTD categories and analyzed the effectiveness of each technique on the cloud through simulation. In~\cite{nhlabatsi2018threat}, the authors proposed a threat-specific risk assessment for the cloud which allows the security administrator of the cloud provider to make decisions for selecting mitigation strategies in order to protect the computing resources of the clients based on the specific security needs and various threats. However, the process of security analysis is performed by the cloud providers. In our proposed approach, we design a secure automated protocol as a platform enabling the IT security experts of the enterprises to analyze the security of their infrastructures on the cloud through their servers from outside the cloud which can offer more flexibility and trust to the organisations.

Moreover, most of the proposed MTD techniques for the cloud lack real cloud implementation, and they are mostly theoretical and simulation-based. Furthermore, the existing approaches only proposed the devising techniques for VM allocation and migrations in order to find the best performance or achieve a security level, but they do not consider the cloud providers constraints, like restriction on deploying VM-LM technique on the cloud by cloud customers.

\section{Conclusions}\label{sec:conclusion}
Cloud security issues are the biggest challenge for enterprises avoiding them to migrate into the cloud. Although the cloud providers consider some security mechanisms, the organizations also need their own security monitoring, analysis, and defensive mechanism to keep their migrated infrastructure secure and safe in the cloud. VM-LM feature of the cloud has been used in many studies as an effective technique that can improve cloud's security. However, most of the cloud providers restrict this feature for their clients. We proposed a framework for organizations migrated into the cloud enabling them: (i) to obtain the security situation awareness of their infrastructures in the cloud. (ii) to plan and select a desirable response strategy such as VM-LM technique to reduce risk and defend against the malicious co-resident VMs. (iii) to securely request a desirable VM-LM strategy to the cloud provider's server for real deployment. 

\bibliographystyle{IEEEtran}
\bibliography{IEEEabrv,MTD}

\end{document}